\documentclass[journal]{IEEEtran}
\makeatletter
\def\ps@headings{%
\def\@oddhead{\mbox{}\scriptsize\rightmark \hfil \thepage}%
\def\@evenhead{\scriptsize\thepage \hfil\leftmark\mbox{}}%
\def\@oddfoot{}%
\def\@evenfoot{}}
\makeatother
\usepackage{latexsym}
\usepackage{amssymb}
\usepackage{epsfig}
\usepackage{amsthm} 
\usepackage[hang]{subfigure}
\usepackage{graphics,color}
\usepackage{amsmath}
\usepackage{algorithm}
\usepackage{algorithmic}
\usepackage{cite}
\usepackage{comment}
\usepackage{url}    
\usepackage{xspace}
\usepackage{psfrag}
\usepackage{alltt}
\usepackage{flushend}

\theoremstyle{definition}

\theoremstyle{definition}

\DeclareMathOperator*{\argmin}{arg\,min}

\IEEEoverridecommandlockouts
\begin{document}
\title{Achieving Congestion Diversity\\ in Multi-hop Wireless Mesh Networks}
\author{ \IEEEauthorblockN{   A. A. Bhorkar\IEEEauthorrefmark{1},  T. Javidi\IEEEauthorrefmark{1}, A. C. Snoeren\IEEEauthorrefmark{2}}                            
 
 \thanks{
 This work was partially supported by the UC Discovery Grant \#com07-10241, Ericsson, Intel Corp., QUALCOMM Inc., Texas Instruments Inc., CWC at UCSD.
A. A. Bhorkar and T. Javidi are with the Department of Electrical and Computer Engineering, University of California San Diego, La Jolla, CA 92093 USA.  A. C. Snoeren  is with the Department of Computer Science, University of California San Diego, La Jolla, CA 92093 USA.
(email: abhorkar@ucsd.edu; tjavidi@ucsd.edu; snoeren@cs.ucsd.edu).  
  }
} 

 \maketitle
\begin{abstract}
This paper reports on the first systematic study of congestion-aware routing algorithms for wireless mesh networks to achieve an improved end-end delay performance.  
In particular, we compare 802.11 compatible implementations of a set of congestion-aware routing protocols against our implementation of state of the art shortest path routing protocol (SRCR).
We implement congestion-aware routing algorithms Backpressure (BP), Enhanced-Backpressure (E-BP) adapted from \cite{Neely06,Neely09} suitably adjusted for 802.11 implementation. We then propose and implement Congestion Diversity Protocol (CDP) adapted from \cite{parul07} recognizing the limitations of BP and E-BP for 802.11-based wireless networks. 
SRCR solely utilizes link qualities, while BP relies on queue differential to route packets. 
 CDP  and E-BP rely on distance metrics which take into account queue backlogs and link qualities in the network. 
E-BP computes its metric by summing the ETX and queue differential, 
 while CDP determines its metric by calculating the least draining time to the destination.

 Our small testbed consisting of twelve 802.11g nodes 
enables us to empirically compare the performance of congestion-aware routing
protocols (BP, E-BP and CDP) against benchmark SRCR. 
For medium to high load UDP traffic, we observe that CDP exhibits significant improvement with respect to
both end-end delay and throughput over other protocols with no  
loss of performance for TCP traffic. Backpressure-based routing algorithms (BP and E-BP) show poorer performance for UDP and TCP traffic. Finally, we carefully study the  effects of the modular approach to congestion-aware routing design in which the MAC layer is left intact. 
\end{abstract}
\section{Introduction}

Traditionally in communication networks, end-end rate adaptation, traffic engineering, and transport layer signaling have been widely developed to 
prevent network congestion. On the other hand, the routing layer is tasked to identify 
shortest-paths to the destination independent of the congestion in the network. In wireless context, variants of shortest-path~\cite{Bicket05,Bicket03,Pei00, padhye04} have been proposed, without modifications  
to the functionalities of the traditional layers, relying various notions distance metric to the destination. A well known example of this approach is SRCR proposed in~\cite{Bicket05}, whose distance metric is based on the number of transmissions required to relay a packet. However, such a shortest path approach to routing traffic falls short in providing acceptable service in
wireless networks as the traffic demand approaches the network capacity and when UDP flows constitute a significant proportion of the traffic.
More precisely, shortest path solutions are susceptible to an underutilization of 
path diversity resulting in increased delay, congestion, buffer overflows, and queue instability.  In contrast,  going back to the seminal work on Backpressure (BP) routing of Tassiulas and Ephremides \cite{TassEph92},  a slew of theoretical and simulation-based studies~\cite{Neely06, Neely09,Mohammad10,Ying11, Bhorkar12} have argued for 
congestion-aware routing protocols: protocols that route packets using only neighbors' congestion levels~\cite{Neely06} or overall congestion along the path~\cite{Neely09,Mohammad10,Ying11} to the destination. 

This paper provides a comprehensive approach to the design, implementation and experimental evaluation of the congestion-aware routing against state of art routing  SRCR  in multi-hop 802.11-based (WiFi-based) wireless networks.  The salient feature of our approach is our 
equal treatment of theory and experimentation in the design of congestion-aware routing algorithms. The design and the choice of the routing protocols in this study are inspired by the theoretical studies in wireless networks~\cite{TassEph92,Neely06, Neely09,Mohammad10}. We have, however, refrained from a redesign of the network at all layers and functionalities as suggested by these studies. Instead, we have devised a solution on a testbed consisting on commercially available 802.11-based wireless radios to shed light on the implications of incorporating the congestion information at the routing layer. 
More precisely, we have restricted our focus to 1) a low overhead implementation of the proposed protocols in the literature, and 2) a modular solution, where only the functionalities of the routing layer have been modified, leaving the physical (PHY) and the media access control (MAC) layers untouched. This pragmatic approach has allowed us to test the basic promise of the congestion-aware routing including and beyond backpressure. This work provides with the first study which carefully investigates the advantages and pitfalls of backpressure routing and a novel design for end-end delay improvements on a network consisting of inexpensive commercially available components.
Secondly, our modular approach allows us to investigate the advantages of congestion-aware routing at the routing layer
isolated from the benefits of generalized scheduling \cite{TassEph92} or receiver diversity gain\cite{Lott06}.

\subsection{Overview of Results}
In contrast to the shortest path routing, backpressure-based routing uses differential backlogs at the nodes to make routing decisions. In other words, under SRCR~\cite{Bicket05} packets are routed along a path with minimum number of transmission attempts (ETX),  BP ignores any such measure of distance to the destination and instead at each node selects the neighbor with the most negative differential backlogs (in the absence of any such neighbor, the node retains the packet). The BP algorithm, effectively balances the queue backlogs at every location and provides with an important theoretical guarantee of throughput optimality (bounded expected delay for all stabilizable traffic). However,  in contrast to SRCR, BP ignores the
distance of the potential forwarders to the destination leading to a worse performance particularly at the low traffic \cite{parul07}. To address the shortcomings of the above solutions, many throughput optimal algorithms have been proposed
  that attempt to integrate the two approaches ~\cite{Neely09,Mohammad10,Ying11} in the opportunistic context.\footnote{In opportunistic setting, the next hop is chosen after the receiving nodes of a packet are known at the transmitter.} 

In this work, we provide a comprehensive evaluation of various congestion-aware solutions in the context of a 802.11-based networks with minimal modifications to the MAC or PHY operations, the only such study of which we are aware. First, we provide a low overhead practical implementation of backpressure-based algorithms BP and Enhanced-BP(E-BP) motivated by the theoretical studies proposed in 
the opportunistic routing context~\cite{Neely06, Neely09}.  In particular, we modify DIVBAR\cite{Neely06} and EDIVBAR~\cite{Neely09}, which rely on the receiver diversity and arrive at designs of BP and E-BP consistent with the widely used MAC of 802.11 wireless cards.  Motivated by a  theoretical and simulation study in the context of opportunistic routing context~\cite{Mohammad10,Bhorkar12}, we also propose a novel alternative Congestion Diversity Protocol (CDP) realizing the drawbacks of the backpressure-based algorithms. The design of these algorithms is unified via a 
class of asynchronous distributed distance-vector routing algorithm similar to the distributed Bellman Ford computations of the ETX. More specifically, we design congestion measures of BP, E-BP and CDP, whose distributed computations are based on an asynchronous exchange of congestion information amongst neighboring nodes. 
These congestion measures are then used to determine the next hop for each packet transmission. This unified approach enables a fair comparison amounts the candidates of interest. 

The main contributions of our work include: 
\begin{itemize}
\item  
We have implemented and studied the candidate routing algorithms: SRCR, BP, E-BP and CDP.
We have taken a pragmatic approach by implementing these solutions on the existing off-the-shelf embedded Alix nodes \cite{Alix} at the routing layer, leaving rest of the radio functionalities untouched. 
\begin{itemize}
\item For TCP traffic, where the transport layer responds to the
  congestion, CDP shows a comparable performance with respect to SRCR. 
  In contrast, BP improves the total throughput by significantly increasing the throughput for short
flows at the cost of severe disruption to long flows and hence fairness. 
\item For UDP traffic, at low traffic loads, CDP performs identical to SRCR, while BP and E-BP show poor performance due to occasional busrtiness and random walk effects in the network.\footnote{The desirable performance of SRCR under low traffic indicates the sufficiency of shortest path solutions in a network with a significant gap between link capacities and ingress traffic, as it is the case with wired networks.} 
\item For UDP traffic in medium to high loads, CDP reduces delay,
  decreases packet drop rate, and increases throughput 
in comparison with BP, E-BP (congestion-aware schemes) and SRCR (congestion blind scheme) in at least 60\% of the scenarios.
\end{itemize}
\item  As a by-product of the modularity of our approach, we identify intra-flow and inter-path interference as the main potential drawback 
of the congestion-aware routing. 
\begin{itemize}
\item 
We provide pathological examples where the intra-flow and inter-path interference significantly over-shadows the benefits of congestion awareness and path diversity. Furthermore, these  pathological behaviors are shown to be inevitable side effects of any modular approach to routing in which MAC layer is kept intact. 
\item In real networks consisting of low rate background traffic,    
a modular implementation is sufficient to capture the benefits of congestion diversity. 
In other words, we show that the pathological examples are not likely to arise in practically relevant situations. 
\end{itemize}
\end{itemize}
 \subsection{Related Work}
We close this section with a note on the related work. While,  much experimental research has shown the value of using the differential backlog information in wireless networks for scheduling and rate allocation at the MAC layer \cite{Walrand}, 
\cite{Hull04mitigatingcongestion}, \cite{Rafael} and the transport layer\cite{Rhee09}, there are very few experimental studies that have dealt with a practical implementation of backpressure as a routing solution with commercially available radios. For example, \cite{Rafael} incorporates the backlog information at all layers making it hard to come to conclusions 
regarding the value of the congestion information at the routing layer. Recently, the authors in \cite{Krishnamachari09} have proposed  Backpressure Collection Protocol (BCP) for sensor networks on top on 802.15.4, to enhance data collection at a single sink node. The BCP implementation, however, requires an impractical LIFO discipline at the MAC layer leading to significant reordering.

In literature,  many heuristic load balancing and multipath routing solutions  \cite{MSR00,SMP01,Path02,PAgrawal06,Das01, Bhorkar12,Mohammad10}  have been proposed for wireless networks in manners that are reminiscence of our approach. 
 To the best of our knowledge, these approaches are studied in terms of simulations and no real implementation is known with the exception of Horizon~\cite{MSR08}.  Horizon~\cite{MSR08} takes load balancing decisions
over two disjoint routing paths (generated separately by a costly link state routing protocol) taking into account queue
backlog information to enhance TCP performance. 

The remainder of this paper is organized as follows.  
Section \ref{routingcongestion} introduces the routing algorithms,
CDP, BP, E-BP and SRCR. In Section \ref{implementation}, we discuss various implementation
issues for these protocols. Section \ref{experiment} provides the performance results for UDP and TCP traffic. In Section \ref{roleinterfereces}, we analyse the performance of the congestion-aware routing algorithms closely with respect to  and determine 
the causes of performance gains and losses by analysing various scenarios.  Finally, we conclude the paper in Section \ref{conclusion}. 

\section{Routing with Congestion Diversity}
\label{routingcongestion}
 
In this section, we start with our 802.11-compatible design for the congestion-aware routing algorithms BP, E-BP and CDP. 
Next, we summarize the exiting design and our implementation for state of the art protocol SRCR.
    
 \subsection {Congestion-aware approach}
The guiding principle of congestion-aware routing has been congestion avoidance in the network
taking into account the queue backlog information and the link qualities between each pair of nodes. 

BP, E-BP and CDP take routing decisions by exchanging a time-varying 
congestion-aware metric, referred to as the {\it{congestion measure}}. 
 For a set of nodes $\Omega$, we denote the congestion measures for destination $d \in \Omega$ at node $n \in \Omega$ as $V_{X,t}^d(k)$, where
 $X$ is the protocol of interest in the set $\{\mbox{BP, E-BP, CDP}\}$. 
In practice $V_{X,t}^d(k)$ is only known at node $n$ via periodic updates received from node $k$. 
Let $\tilde{V}_{X,t}^{(n,d)}(k)$ be the latest congestion measure advertised by neighbor $k$ and received at node $n$. 
Based on the received congestion-measure $\tilde{V}_{X,t}^{(n,d)}(k)$, each node $n$ in the network updates its routing table.  In particular, the routing table
determines the next-hop $K_{X,t}^{(n,d)}$ for a packet at node $n$ destined for node $d$. After each successfully acknowledged transmission, the routing
responsibility is then transferred to the next hop. The congestion-aware algorithm also performs a flow selection among the packets associated with different destinations using virtual queue mechanism at layer-2. In particular, node $n$ selects the packet destined for node $m_{X,t}(n)$ among available packets and transmits the packet to the PHY layer. Table~\ref{notations} provides notations used in the description of these algorithms.

\begin{figure}[t]
\centering
\includegraphics[width=\columnwidth]{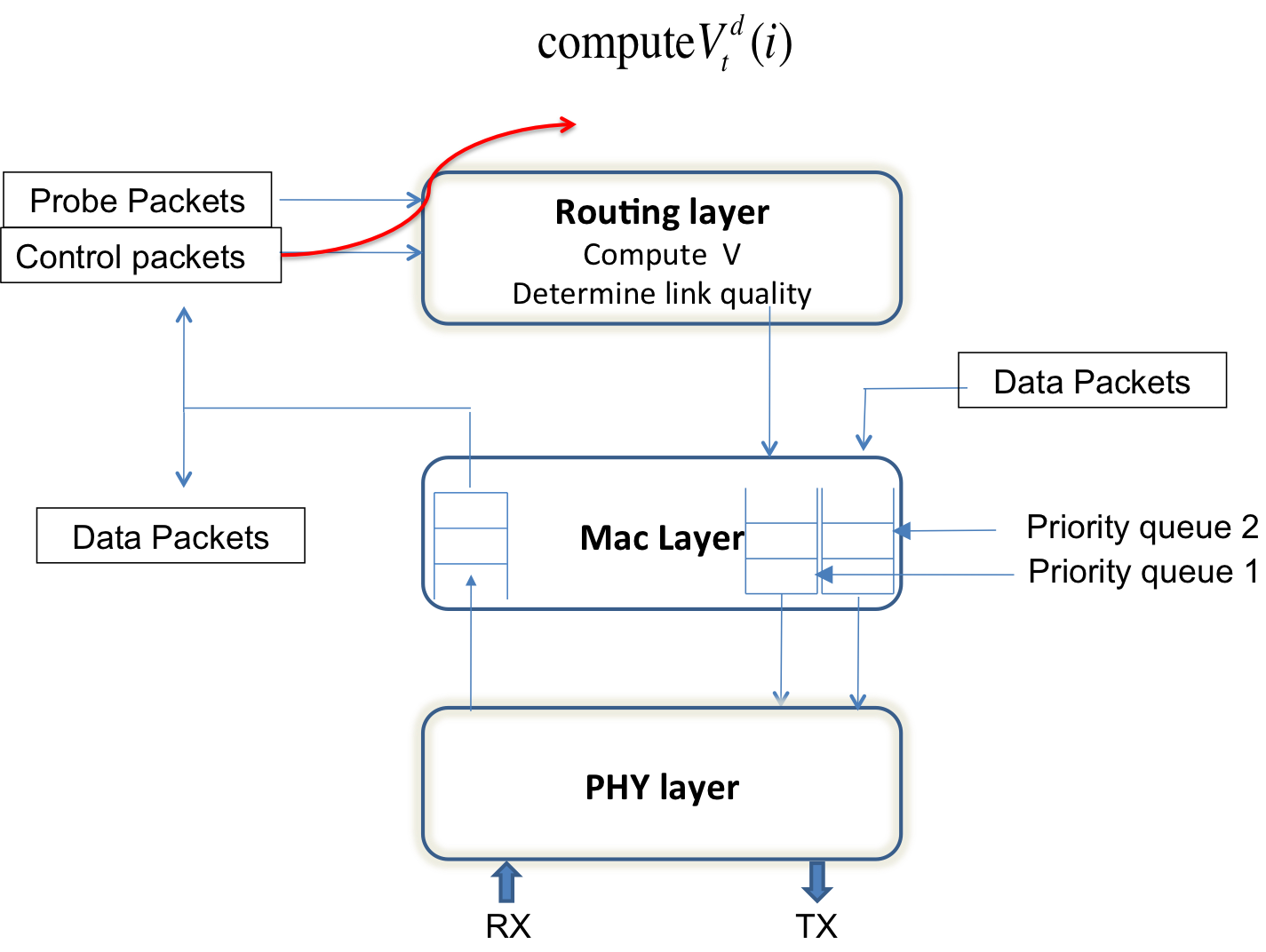}%
\caption{Design of congestion-aware routing algorithms. }%
\label{operation}%
\end{figure}

The design of these congestion-aware algorithms rely on a routing table at each node to determine the next best hop. 
The routing table at node $n$ consists of a list of neighbors $\mathcal{N}(n)$,  
a structure consisting of estimated
congestion measures $\tilde{V}_{X,t}^{(n,d)}(k)$ for all neighbors  $k \in \mathcal{N}(n)$ associated with different destinations, and 
the best next hop vector $\{ K_{X,t}^{(n,d)}\}_{d \in \Omega}$. 
Node $n$ periodically 
advertises the entries of the its computed congestion measures to its neighbors at intervals of $T$ seconds using control packets. Thus, the periodic computation and communication of congestion-measures   
propagates routing information across the neighbors. The sequence of operations performed are shown in Figure~\ref{operation}.

BP, E-BP and CDP have different notions of measuring the effective congestion in the network 
and thus determining next hop selections. Next, we detail the computations performed at each node to determine the congestion measures and next hops for BP, E-BP and CDP respectively.

\begin{table}%
\centering
\caption{Notations used in the description of the algorithms}
\begin{tabular}{|c|l|}
\hline 
Symbol  & Definition \\
\hline
 $\mathcal{N}(n)$ & Neighbours of node $n$ \\  
    & \\
     $W(n,k)$ & Transmission time on the link from node $n$ to $k$\\
            &  \\    
    ${q}_t^d(n)$ & Queue-length at node $n$ destined for $d$ at time $t$\\
            &  \\  
  $ETX^{(k,d)}$ & ETX from node $k$ to destination $d$\\
            &  \\            
 $V^d_{X,t}(n)$ &  Congestion measure computed for protocol $X$ at node $n$\\
           & $X \in \{\mbox{BP, E-BP, CDP, SRCR}\}$\\
          & \\
 
 $\tilde{V}_{X,t}^{(n,d)}(k)$ & Latest congestion-measure obtained at $n$ from node $k$\\
            &  \\          
  $K_{X,t}^{(n,d)}(t)$ & Selected relay by protocol $X$ at node $n$ \\
          & \\   
      $m_{X,t}(n)$ &  Packet with destination $m_{X,t}(n)$ is selected \\
          &  by protocol $X$ at node $n$ \\
          & \\                          
            \hline
\end{tabular}
\label{notations}
\end{table}

 \subsubsection{Backpressure Protocol (BP)}
 For BP, the congestion measure is simply the queue backlog information. In particular, each node advertises its current 
 queue-backlog information for each destination as a congestion measure in the control packet. 
 The congestion measure ${V}_{BP,t}^d(n)$ for node $n$, $n \ne d$ is given as 
${V}_{BP,t}^d(n) =  q_t^d(n)$.
 Thus, effectively, the estimated congestion measures $\tilde{V}_{BP,t}^{(n,d)}(k)$ at node $n$ denotes the latest queue length information at  its neighbor $k \in \mathcal{N}(n)$.     

 The BP then selects the next hop based on a weighted differential backlog. 
For any destination $d$, BP chooses the next hop $K_{BP,t}^{(n,d)}$, such that: 
    \begin{equation}
    \label{kbp}
     K_{BP,t}^{(n,d)}   =  \argmin_{k \in {\mathcal{N}}(n)} \frac{1}{W(n,k)} \left (\tilde{V}_{BP,t}^{(n,d)}(k) - V_{BP,t}^{d}(n) \right ). 
    \end{equation}

BP performs a flow selection by selecting packets with destination $m_t(n)$ among all possible virtual queues for all possible destinations
 with minimum queue differential such that: 
  \begin{equation}
     \label{commodity}
      m_{BP,t}(n)   =  \argmin_{d} \min_{k \in {\mathcal{N}}(n)} \frac{1}{W(n,k)}\left (\tilde{V}_{BP,t}^{(n,d)}(k) - V_{BP,t}^{d}(n) \right ).  
    \end{equation}               
              
The original backpressure \cite{Neely06} assumes a globally synchronized time-slotted MAC protocol as well as a controller that computes and  schedules the nodes in centralized manner. 
Note that our implementation of BP is an approximate variant of the original backpressure proposed in \cite{Neely06} adjusted 
for the distributed implementation on 802.11-based networks.

\subsubsection{Enhanced Backpressure Protocol (E-BP)}
  E-BP is a variant of BP, where along with the queue information, the ETX metric is used for path selection. E-BP, similar to BP, uses queue backlog information as 
 the congestion measure and ${V}_{E-BP,t}^d(n) =  q_t^d(n)$.  
 Furthermore, for a packet destined for $d$, E-BP chooses its next hop $K_{E-BP,t}^{(n,d)}$ such that: 
        \begin{eqnarray}
         \nonumber
     K_{E-BP,t}^{(n,d)}  & = & \argmin_{k \in {\mathcal{N}}(n)} \Big \{ ETX^{(k,d)}    \\
       &+ & \frac{1}{W(n,k)} \left (\tilde{V}_{E-BP,t}^{(n,d)}(k) - V_{E-BP,t}^{d}(n) \right ) \Big \} ,   
    \end{eqnarray}   
 where $ETX^{(k,d)}$ is the minimum  transmission time from node $k$ to destination $d$ defined as              
     \begin{equation}
     \label{csrcr}
     ETX^{(k,d)}   =  \min_j \Big \{ ETX^{(j,d)} + W{(k,j)} \Big \} . 
    \end{equation}

  Finally, the packet with destination $m_{E-BP,t}(n)$ is selected among all possible destination packets which minimizes sum of queue differential and ETX 
  such that: 
  \begin{eqnarray}
     \label{commodityE-BP}
     \nonumber
      m_{E-BP,t}(n)  & = & \argmin_{d}  \Big \{ \min_{k \in {\mathcal{N}}(n)}  ETX^{(k,d)} + \frac{1}{W(n,k)}  \\
        & \times & (\tilde{V}_{E-BP,t}^{(n,d)}(k) - V_{BP,t}^{d}(n))    \Big \}. 
    \end{eqnarray}
 
Note that, for E-BP, the control packet needs extra overhead to compute the ETX (see Section \ref{srcr}) along with the transmission of the congestion-measure.  
 
 \subsubsection{Congestion Diversity Protocol (CDP)}
 
The congestion measure for CDP is the aggregate sum of the local draining time at the node $n$ and the draining time from its next hop till the destination. 
In CDP, when relaying packets destined for node $d$, node $n$ selects the targeted receiver $K_{CDP,t}^{(n,d)}$ to minimize the packet's delivery time, i.e. 
\begin{eqnarray}
\label{updatenexthop} 
	K_{CDP,t}^{(n,d)} & = & \argmin_{k \in {\mathcal{N}}(n)}
	 \left \{ {W(n,k)} + \tilde{V}^{(n,d)}_{CDP,t}(k) \right \}.  
\end{eqnarray}

CDP takes a simplistic approach in flow selection. In particular, it does not use virtual queues for different destinations, rather CDP uses 
a FIFO discipline at layer-2 for all packets. 

Assuming a FIFO discipline at layer-2, we proceed to describe the computations of congestion measure for CDP. 
The congestion measure associated with node $n$ for a destination $d$ at time $t$ 
is the aggregate sum of the local draining time at node $n$ and the estimated draining time from its next hop, $\tilde{V}_{CDP,t}^{(n,d)}(K_{CDP,t}^{(n,d)})$. 
  The local draining time for a packet destined for $d$ arriving at $n$ at time $t$ is equal
to the duration of the time spent draining the packets that arrived earlier plus its own packet delivery time.
In other words, if $q_t^j(n)$ is the number of packets  destined for $j$ queued at node $n$ at time $t$, the local draining time is equal to 
\begin{eqnarray}
\label{drain2}
\sum_{j \in \Omega} {q_t^j(n)}{W(n,K_{CDP,t}^{(n,j)})} + W(n,K_{CDP,t}^{(n,d)}) . 
\end{eqnarray}
The congestion measure for node $n$, $n \ne d$ is then given as 
\begin{eqnarray}
\nonumber
V_{CDP,t}^d(n) &= &  {W(n,K_{CDP,t}^{(n,d)})} + \sum_{j \in \Omega} {q_t^j(n)}{W(n,K_{CDP,t}^{(n,j)})}  \\
\label{VnCDP}
&&   +   \tilde{V}_{CDP,t}^{(n,d)}(K_{CDP,t}^{(n,d)}). 
\end{eqnarray}
Thus, in CDP, the congestion measures are computed in a fashion similar to distributed Bellman-Ford computations~\cite{Lott06}.  

 \subsection {Congestion-unaware approach: SRCR}
 \label{srcr}
 For the sake of completeness, we describe the design of congestion unaware, shortest path routing protocol SRCR.  This implementation acts as a benchmark for the comparison purposes with respect to the congestion-aware routing. 
SRCR  proposed in \cite{Bicket05} uses the ETX metric when routing the packet considering only the link quality information at the nodes.
The ETX to reach the destination is computed using the transmission duration $W(i,j)$ between each pair of nodes $i$ and $j$. 
 Specifically, 
for a packet destined for node $d$, the next hop $K_{SRCR,t}^{(n,d)}$ is chosen such that
      \begin{equation}
      \label{ksrcr}
     K_{SRCR,t}^{(n,d)}   =  \argmin_k ETX^{(k,d)} +{W{(n,k)}}, 
    \end{equation}
 where $ETX^{(k,d)}$ is the minimum  transmission time from node $k$ to destination $d$ computed as (\ref{csrcr}).              
We use the distributed architecture of CDP for the calculation of ETX metric by taking $q^d_t(n) = 1$ in the calculation of $V^d_{CDP,t}(n)$.

In the next section, we discuss the practical issues associated with computation of
the congestion measures for these congestion-aware routing algorithms.
Furthermore, we propose practical implementations and heuristics.

\section{Implementation Details}
\label{implementation}

In this section, we provide the elements of the congestion aware routing responsible for the 
computation of the congestion measure including reliability of control packets, link quality estimation, neighbor discovery, flow selection, and avoidance of loops while routing.

\subsection{Control Packet Reliability}

An important component of the implementation is the exchange of congestion-measures using control packets. In particular,
BP, E-BP and CDP depend on a reliable, frequent, and timely delivery of the control packets.  
As documented in \cite{Shaikh00routingstability}, the loss of control packets can be the cause of instability in  
many well known routing algorithms. Thus, it is important that our implementation
  ensures a high reliability for the delivery of control packets.
In our implementation, we have 
taken advantage of the priority-based 802.11e to implement this component of the control plane. In other words, 
the MadWifi scheduler assigns highest strict priority to the control packets.    
It reduces the probability of control packet drop at the MAC layer and also ensures their timely delivery. In the context of 802.11e, we  utilize two of the four  priority queues: data packets are assigned to the lower priority queue (WME-AC-BK) and control packets are assigned to the  
higher priority queue (WME-AC-VO) \cite{madwifi}.
 In our implementation, the scheduler assigns a 
sufficiently reliable and low PHY rate (11 Mbps in our testbed) for the control packets. These design choices ensure high reliability and speedy delivery of the control packets. 

{
\subsection{Link quality estimation}
\label{LPP}
 The computations given by (\ref{kbp})-(\ref{ksrcr}) utilize the transmission time $W(i,j)$ for each pair of nodes $i$, $j$. 
 In this work, we measure the transmission time $W(i,j)$ by taking the difference between the
  instant when a packet enters the hardware at node $i$ and when the acknowledgement is received from node $j$ at 
  node $i$.\footnote{Different hardware-independent proxies for transmission times exists in the literature and these can be of interest for easy implementation. For example, link success probabilities or the number of retries \cite{samplerate05} can be used to obtain approximate transmission times.}  
  
The transmission time $W(i,j)$ for a packet from node $i$ transmitted to node $j$ can be written as the interval 
between the reception of an acknowledgement (ACK) at node $i$  from node $j$ and the transmission of the packet from node i's interface. 
The driver and the hardware may provide functionalities to accurately measure the transmission time (up to the accuracy of the operating system scheduling delay). Specifically, in the context of atheros cards, we will detail the computation of $W(i,j)$. 
For atheros-based wireless interfaces and the MadWifi driver used in our experiments, 
we can easily determine the instant of time when $i$ receives an ACK from $j$.  
Now, we are left to determine the time instant at which packet enters the interface. 
Unfortunately, the MadWifi driver does not provide the time of entry at the interface directly. Instead, the driver provides
information on when the packet enters the queue at the MAC layer. 
With such limited information at hand, we devise the following algorithm to determine 
drain time. Timestamp $T_1$ when the packet enters the MAC queue, timestamp $T_2$ when the previous packet exits the interface and 
timestamp $T_3$ when the current packet exists the interface. The transmission time is then given as $T_3 - \max(T1,T2)$. 
 
At the system level, we combine the link quality measurements  using actual data packets at the nodes  (passive probing) with  dedicated probe
packets transmitted to each neighbor when a node does not engage in data transmission (active probing). 
These estimates are combined using a weighted average.

{

\subsection{Neighbor Discovery}
Each node needs to maintain information on the cost vectors along with the link quality information for all of its neighbors to efficiently route the packets. 
In order to reduce the overhead, we restrict the set of neighbors by using a sufficiently reliable link.
Specifically, the neighbors are defined so that the link success probability for each neighbor is above a  threshold $\gamma$. 
Defining neighbors using a delivery ratio eliminates any artifacts due to external interference. 
Out of the many possibilities to implement this procedure, we have used dedicated probe packets to obtain the relevant information.  

\subsection{Loop Avoidance}
\label{loop}
Unless carefully designed, distributed computations of any time-varying distance vector routing algorithm are likely to suffer from the classical problem of counting to infinity \cite{Moss82}. Looping can result 
in large delays, increased interference and loss of packets. 
 The problem is most acute when 
there is a sudden burst of traffic,\footnote{Similar to the broken link scenario in a typical distance vector routing.} 
resulting in a transient build-up of queue and an overestimation of the quality of the route.
Such transient effects can be severe due to the resulting slow exchange of control packets.

To address this issue, in case of CDP, 
we utilize Split-horizon with Poison reverse solution \cite{Split00} to avoid loops.
CDP uses  control packet information received from neighbors to gather appropriate information for the split horizon implementation. 
In Split-horizon with poison reverse, a node advertises routes as unreachable to the node through which they were learned. 
Note that, we apply loop avoidance methods only for CDP, while BP and E-BP are left in their original form in \cite{Neely09}.  
\subsection{Flow Selection}
In CDP, we utilize FIFO discipline at layer 2 and we do not employ any flow selection at the packet level. 
In BP and E-BP, we  approximate a packet level flow selection proposed in  \cite{Neely06,Neely09}. 
 The  original backpressure algorithm proposed in \cite{Neely06} proposes a 
  scheduler to choose the destination queue to be served according to to (\ref{commodityE-BP}).  
   In order to implement  priority scheduling, we utilize the 802.11e-based priority scheduler~\cite{Alix}  
    and the highest destination packet is assigned to the higher priority hardware queue. 
    
\section{The Experimental Setup}
\label{experiment}

\begin{figure}[!h]
\centering
\includegraphics[width=0.8\columnwidth]{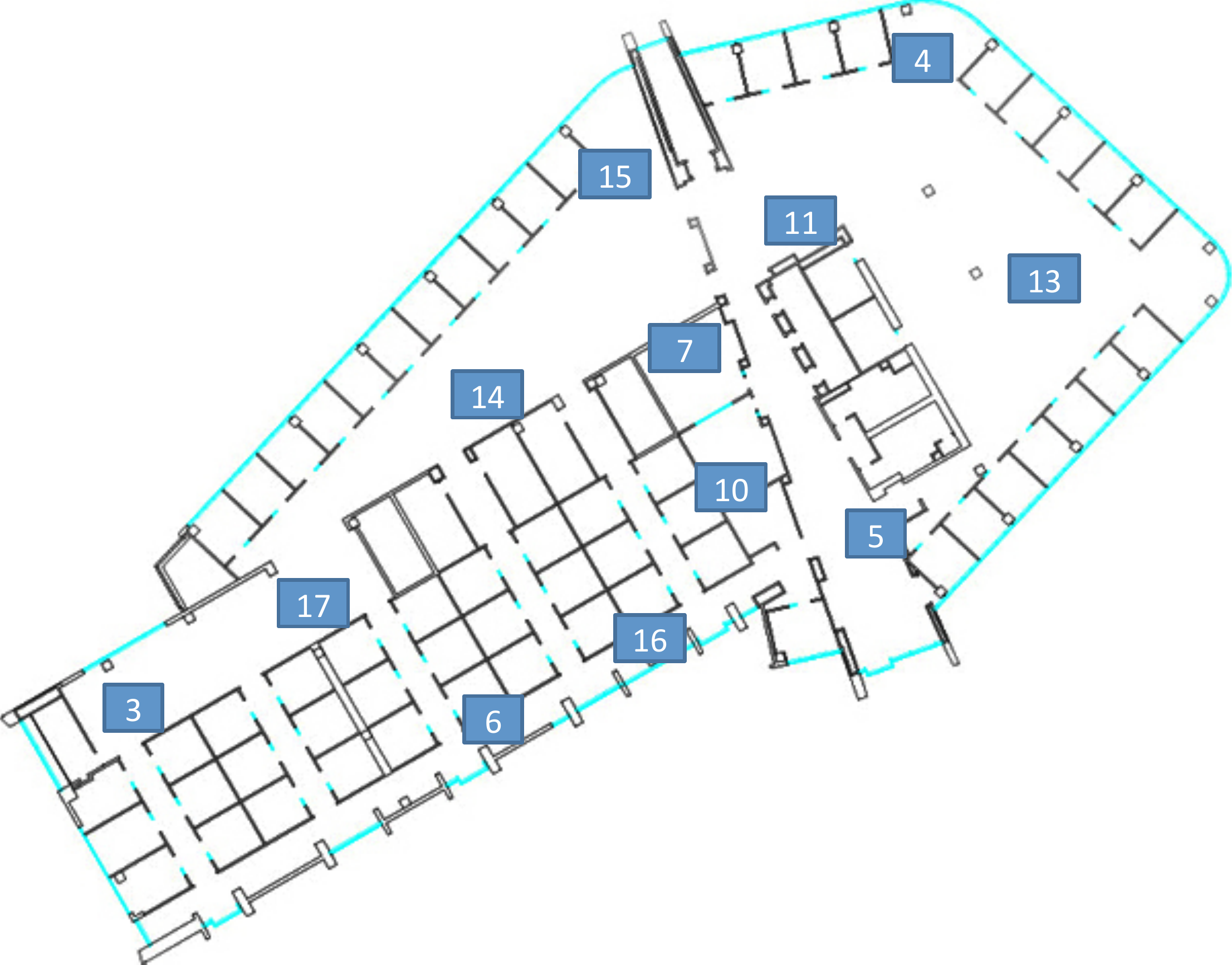}%
\caption{Testbed: Node locations}%
\label{cdpholes2}%
\end{figure}

Our testbed consists of 12 wireless Alix \cite{Alix} nodes with 512 MB
RAM and a 500 MHz processor on Linux version 2.6.22.  The nodes placed as shown in Figure \ref{cdpholes2}, are distributed in
Atkinson Hall, UCSD, in 
about 215,000 sq. ft. of space. Each node is
equipped with an Atheros-based 802.11 a/b/g wireless interface (AR 5213)
connected with omni directional antennas. All nodes are connected to
Ethernet ports for maintenance and data collection.  All nodes are configured in the
802.11g ad-hoc mode with RTS/CTS disabled and the transmission power is set to
13 dBm. In addition to human inhabitants, the building contains hundreds of workstations and a large variety of electronics operating in the same 2.4 GHz unlicensed frequency band as 802.11, resulting in a highly variable channel quality in different portions of the building and during different times of the day. For consistent data, we performed our experiments during the night when the variability of the channel is least.

All of the above routing algorithms have been implemented in user space with appropriate calls to the
MadWifi driver, which is supported by the Linux kernel (2.6.22
onwards). These algorithms perform queuing and scheduling on every packet being
transmitted or received by the driver. 
We have used a transmission rate of 11 Mbps for the control packets while the data packets are sent at 48 Mbps. The packet size for data packets is 512 bytes. For each algorithm, each iteration of traffic generation is executed for 180 seconds. 

We study the choice of the parameters in the subsequent analysis. Specifically, we need to set the following parameters:  control packet interval $T$, probe interval duration, and  neighbor probability threshold $\gamma$. 

 {\it{1) Choice of control packet interval $T$}:}
In our setup, we transmit control packets periodically at intervals of 200 ms. Each control packet of roughly 200 bytes is broadcasted at a rate of 11 Mbps. 

We need to tradeoff the overhead of the control packets and the need to obtain  
the accurate congestion measures from neighbours. 
Since broadcast packets do not undergo a backoff mechanism, the broadcasting cost for control packets is negligible compared to data transmission.
In order to study the interference effects of control packets on the performance, we vary the control packet transmission interval for SRCR and compare the performance with various intervals of 50, 100, 200, 300 and 500 ms. 
 We observe that the performance does not vary for intervals greater than 200 ms, which implies that the control packets do not have a significant effect on the performance. 

 {2) \it{Choice of link probe interval}:}
 In our setup, we transmit probe packets at regular intervals to probe the channel and learn the link quality. 
The choice of probing interval should trade-off the added overhead with the ability to track channel variations. 
We set the probe interval to 1 second in accordance with the value chosen in~\cite{Kim06onaccurate,Morris05}
consistent with indoor environment's fading parameters. Furthermore, probe packets of length 512 bytes 
are selected to match the data packet size (also set at 512 bytes). 

 {3) \it{Choice of probability threshold $\gamma$}:}
We have chosen $\gamma=0.4$ for trading off link reliability with network connectivity. The decision whether a node is a  neighbor, is based on the initial non-interfering condition of the network.

 With the above setup, we are ready to evaluate 
the performance of congestion-aware routing protocols for TCP and UDP traffic as reported next.

\section{Performance Results}
\label{udpexp}
{ 
In this section, we perform a comparative study of various routing protocols under TCP and UDP 
traffic. 
In our comparative analysis, we investigate 
the following performance measures: 
\begin{enumerate}
\item End-End delay: For $M$ packets, we define the mean delay $D=\frac{1}{M}\sum_{j=1}^{M}(\tau_A^{j} - \tau_D^{j})$, where $\tau_A^{j}$ is the arrival time at the destination and $\tau_D^{j}$ is the departure time for packet $j$ at the source. For TCP traffic, we consider mean delay as mean Round Trip Time (RTT). We are interested in the distribution of per packet delay, e.g. the Cumulative Distribution Function (CDF) of 
$D$ with respect to the random choice of network topology. 
For illustration purposes, we consider a differential 
delay measure which consists of the difference between CDP and the candidate protocol. Specifically, we plot the difference $D_{candidate}-D_{SRCR}$, where $D_{SRCR}$ is the mean delay for CDP and $D_{candiate}$ is the mean delay for the comparative protocol. 
\item Throughput ratio:  The throughput is the number of bytes received at the destination for the duration of the experiment. Again when investigating the CDP performance with respect to the network topology, we use the normalized throughput ratio as a measure of performance, where the normalized throughput ratio is defined as the ratio between the throughput of the candidate protocol versus the SRCR. 
\end{enumerate}    
\subsection{Experiments with TCP}
\label{tcpexp}
In this section, we study the performance of congestion-aware routing algorithms for TCP   
used for reliable communications. We report
 the comparative performance of the candidate routing  protocols under reliable transfer control algorithms 
TCP-Veno~\cite{Veno}  by selecting a configuration of two TCP flows with randomly selected source and destination pairs. 
We do not expect to see significant improvement for congestion-aware routing protocols with respect to SRCR. 
The first reason for the insignificant performance gain is that the current implementations of TCP are non-aggressive. Thus, TCP tries to avoid congestion in the network and thus makes the congestion routing insensitive to TCP connections. 
Secondly, TCP is known to have performance degradation with respect to packet reordering thereby increasing the
packet retries at the transport layer. 
Figure \ref{fig:reordertcp} shows that BP, E-BP and CDP suffer from the reordering of packets, leading 
to a further decrease in performance.

\begin{figure}[ht]
\centering
\subfigure[Differential RTT] {
      \includegraphics[width=0.37\textwidth]{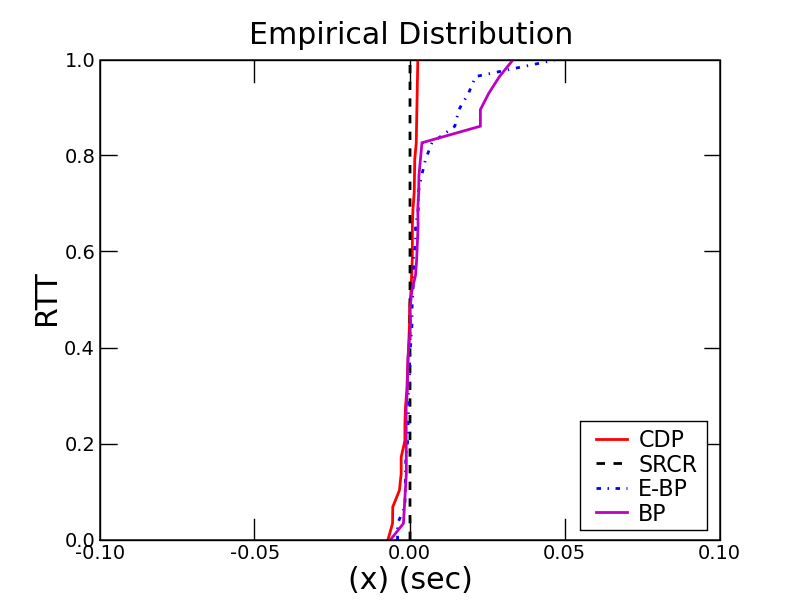}              
 \label{fig:rtt0tcp}
 }
\subfigure[Normalized throughput]{
\includegraphics[width=0.37\textwidth]{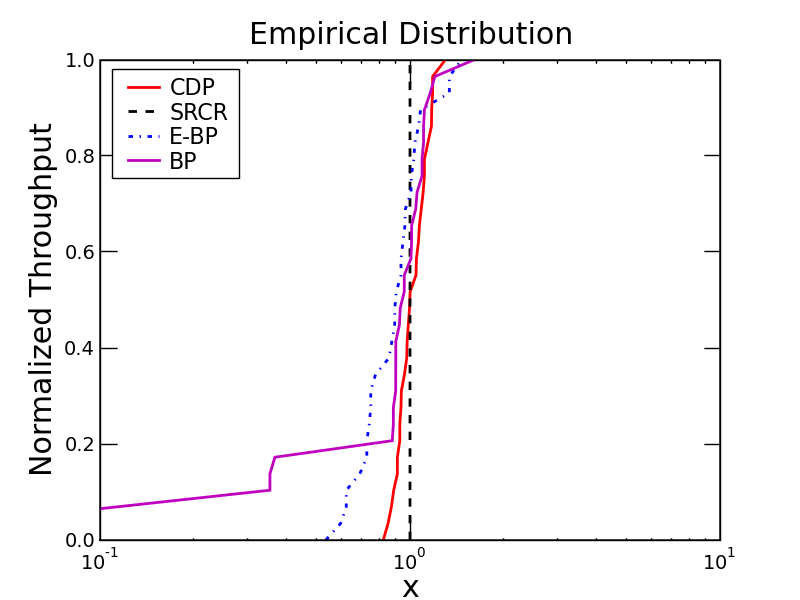}                              
   \label{fig:thru0tcp}
   }
  \caption{Performance for TCP traffic}
\end{figure}

\begin{figure}[ht]
\centering
    \includegraphics[width=0.37\textwidth]{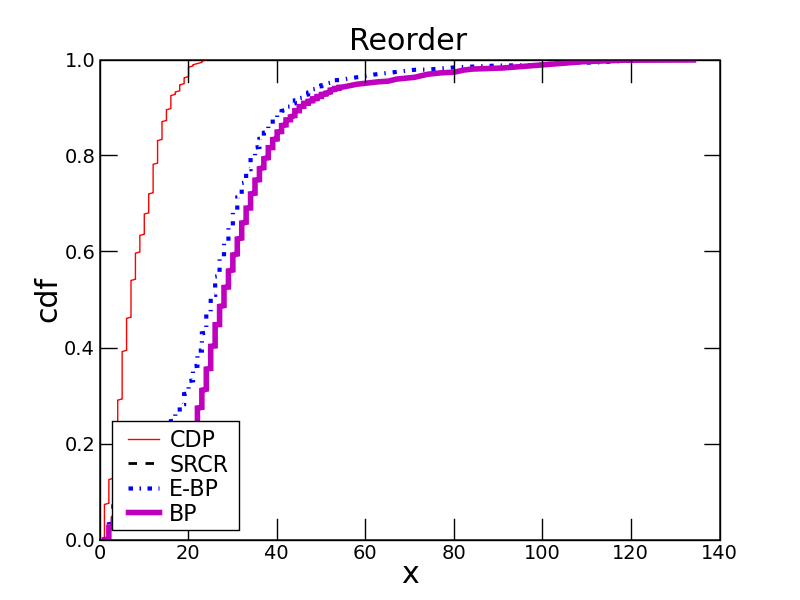}         
\caption{CDF of reordering}
 \label{fig:reordertcp}
 \end{figure}

Figures \ref{fig:rtt0tcp} and \ref{fig:thru0tcp} compare the CDF of the Round Trip Time (RTT) and the normalized throughput for multi-hop flows  for CDP, SRCR, BP and E-BP. 
These set of experiments show that CDP ensures a comparable performance with SRCR for TCP flows. 
For multi-hop flows, the performance of TCP for backpressure-based algorithms suffers from the loops  and ``dead ends'', resulting in timeouts and very low throughput for routes with multiple hops. 

TCP performance results show that CDP is expected to show near equal performance with respect to SRCR while significantly worse performance for BP and E-BP and hence we will not study TCP 
further in this paper. 
Next, we dissect and study the more interesting case of UDP traffic. 

\subsection{Experiments with UDP}	
\label{udponlyexp}
In this section we report on the performance of  BP, E-BP, CDP, and SRCR in our network with two competing randomly selected flow with two sets of source destination pairs. 
We then inject Poisson  traffic at each source node with a randomly selected average load between 0 Mbps and 7 Mbps. 
We repeat the experiment for 100 such configurations. 
We filter out the scenarios which are
unsustainable under any protocol. 
More precisely, we consider configurations for comparison where at least one algorithm delivers 80\% of packets. (15\% of the scenarios were overloaded). 
  Furthermore, uninteresting cases with only single hops are avoided from the analysis (15\% of the scenarios contained only single hops).  
  Figures \ref{rates} plots the offered loads for the remaining configurations in our analysis.
  We  classify the configurations under considerations into two sets: low load  configurations when 
the observed delay for SRCR is sufficiently small (less than 0.1 second) and high load constituting the remaining configurations which are operating near capacity.   
  We report the Cumulative Distribution Function (CDF) of the performance metrics for various protocols  for low and high load scenarios in Figures~\ref{fig:delaylow}-\ref{fig:highgainthru1}.

\begin{figure}[ht]
\centering
    \includegraphics[width=0.8\columnwidth]{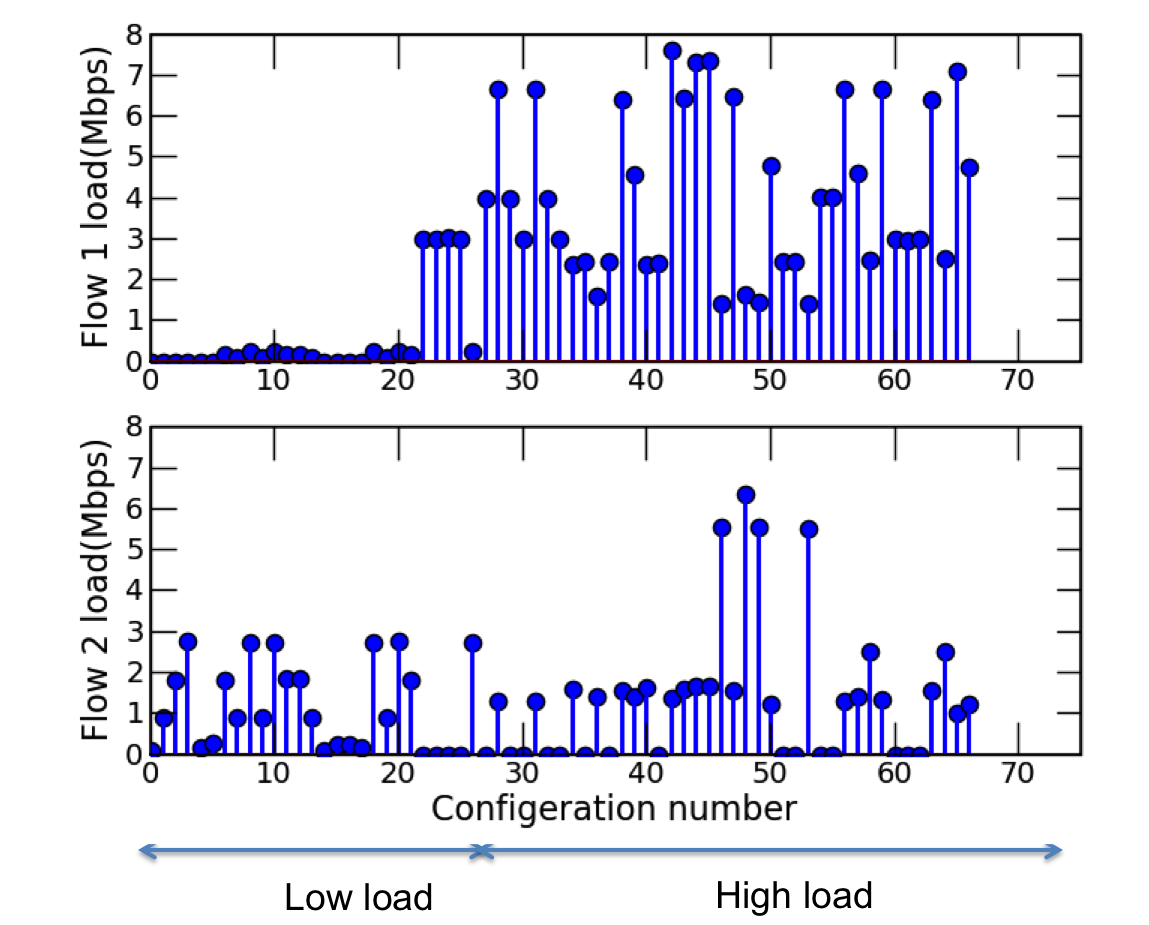}   
\caption{Offered load}
    \label{rates}
\end{figure}

\begin{figure}[ht]
\centering
    \includegraphics[width=0.8\columnwidth]{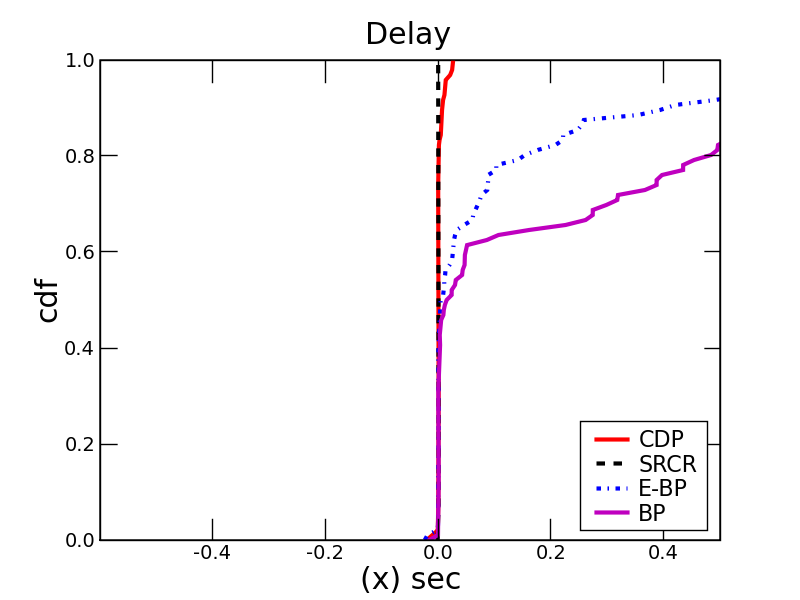}   
\caption{CDF of delay differential for a low load}
    \label{fig:delaylow}
\end{figure}
\begin{figure}[ht]
\centering
    \includegraphics[width=0.8\columnwidth]{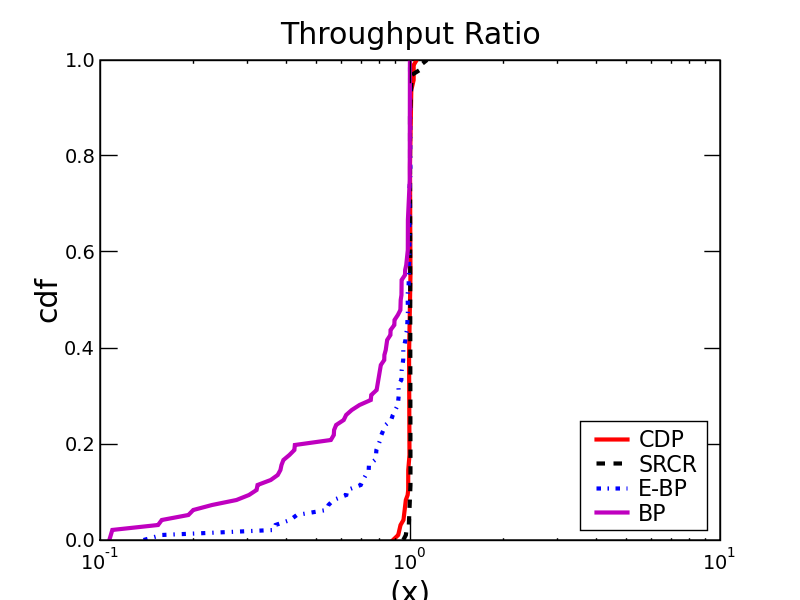}   
\caption{CDF of normalized throughput for low load}
    \label{fig:thrulow}
\end{figure}

In Figures~\ref{fig:delaylow} and \ref{fig:thrulow}, we plot the delay differential and normalized throughput for the candidate protocols under a low load scenario. Here, as we expected, CDP  performs similar to SRCR while significantly performing better than BP and E-BP. 
This is because in absence of congestion the distributed computation in (\ref{VnCDP}) reduces to computation of the ETX, while BP and E-BP reduce to a near random walk in the network. 

Figure \ref{fig:highgaindelay1} compares the CDF of the delay differential, while    
Figure \ref{fig:highgainthru1} shows the CDF of the normalized throughput ratio for high traffic load. Figures~\ref{fig:highgaindelay1}, and \ref{fig:highgainthru1}  show that for about 60\% of the network configurations selected, CDP delivers packets with significantly less delay compared to other protocols\footnote{The candidate protocol performs poorly if the CDF lies to the left of the SRCR}. Figures~\ref{fig:highgaindelay1} and \ref{fig:highgainthru1} also show the possibility of low to moderate performance loss under CDP (compared to SRCR) in 25\% of the scenarios. 
In Section \ref{analysis} we dissect and isolate the sources of loss in these scenarios.

\begin{figure}[ht]
\centering
              \includegraphics[width=0.8\columnwidth]{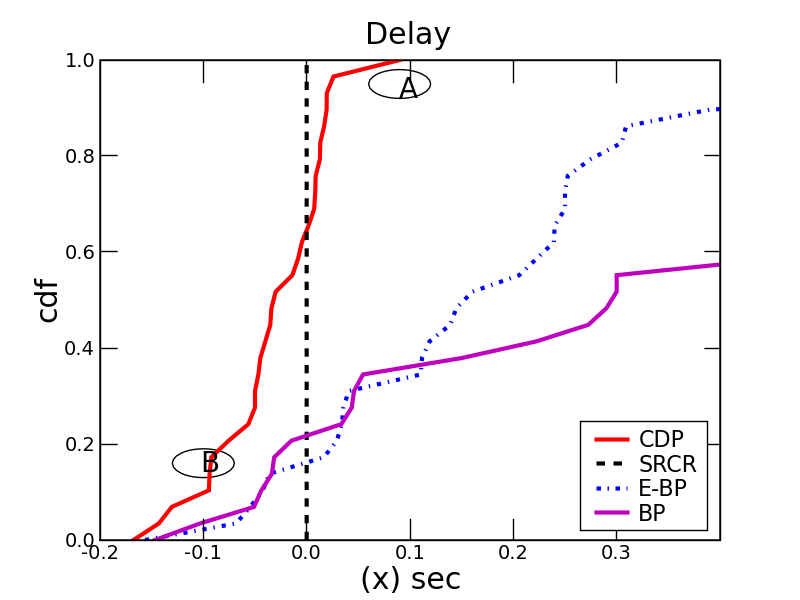}   
\caption{CDF of delay differential for high load}
 \label{fig:highgaindelay1}
\end{figure}

\begin{figure}[ht]
\centering
        \includegraphics[width=0.8\columnwidth]{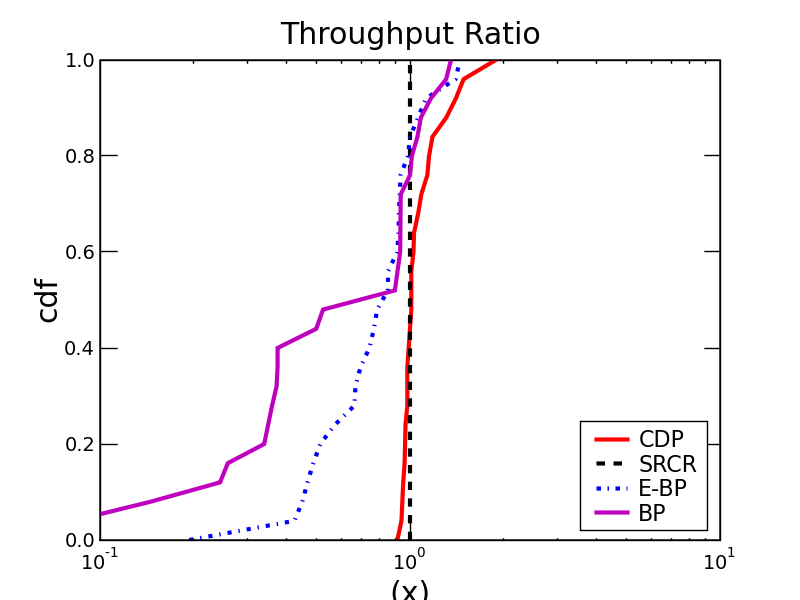}   
\caption{CDF of normalized throughput  for high load}
 \label{fig:highgainthru1}
\end{figure}

\subsection{Congestion awareness : Pros and Cons}	
\label{analysis}
In this section, we investigate further the contributing factors to delay performance for the UDP traffic of Section \ref{udponlyexp}.
To gain insight on the strengths and weaknesses of congestion-aware routing, we consider two examples of 
topologies associated with delay differentials on both ends of the spectrum on the CDF plot in Figure~\ref{fig:highgaindelay1}. 
The performance of these examples with respect to SRCR is marked as point A and point B respectively in Figure~\ref{fig:highgaindelay1}.
The topologies associated with these points A and B are shown in Figure \ref{topo_inter}.
Our first  example topology (point A) consists  
  of one high (4 Mbps) unicast flow between nodes 10-17 and another low load flow (1 Mbps) between
nodes 14-16. In this setting,  the delay performance of CDP  is significantly worse than SRCR.  
 The second topology consists of the same flows 10-17 and 14-16 but with (swapped) flow rates of 1 Mbps and 4 Mbps. Note that this second topology coincides with the case when the CDP significantly outperforms SRCR.

\begin{figure}[th!]
\centering
    \includegraphics[width=0.8\columnwidth]{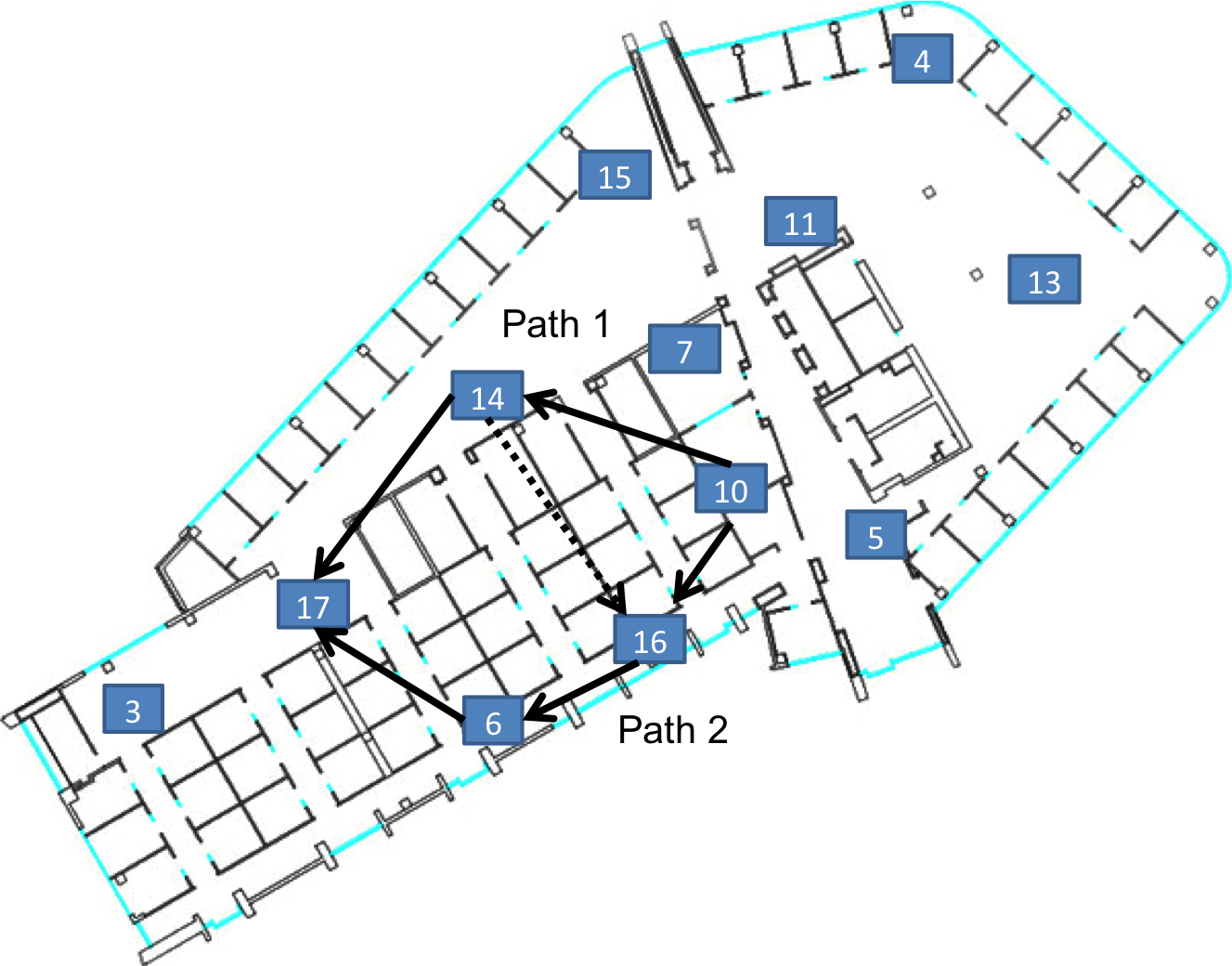}   
    \caption{Topology used to analyse the performance results. Flow 10-17 is studied with and without the presence of flow 14-16. }
\label{topo_inter}
\end{figure}

\subsubsection{Congestion-aware Routing: Cons (point A)}

We study the flow configuration 
at point A consisting of flow 10-17 with high load and 14-16 with low load.  
 Figure \ref{delay1udp_anti} plots the end-end delay for the individual flows. We observe that CDP performs worse  
compared to SRCR, while BP and E-BP show  below par performance. 
 
To understand the sources of loss of performance under congestion-aware routing policies we have  
 illustrated the next hop selections by node 10 in Figure \ref{fig:QueueNextHop_anti} where we plot $K^{(10,17)}$  under the candidate protocols throughout
the duration of the experiments. We observe that SRCR maintains node 14 as the next hop in a static manner while CDP switches its next hop from node 14 to node 16. 
 BP and E-BP forward significant number of packets into nodes 5, 7 and 11 increasing the interference in the network.

Figure \ref{loss_anti} decomposes the loss incurred by each protocol into buffer overflow, retry loss and loop loss.\footnote{Loop loss occurs due to the presence of loops in routes resulting in packet drop if the Time to Live (TTL) value reaches 0.} 
In this example, we observe flow 10-17 suffers a higher queue loss for CDP compared to SRCR. Other losses remain fairly negligible (even though higher for BP and E-BP).

\begin{figure}[!ht]
\centering
    \includegraphics[width=0.8\columnwidth]{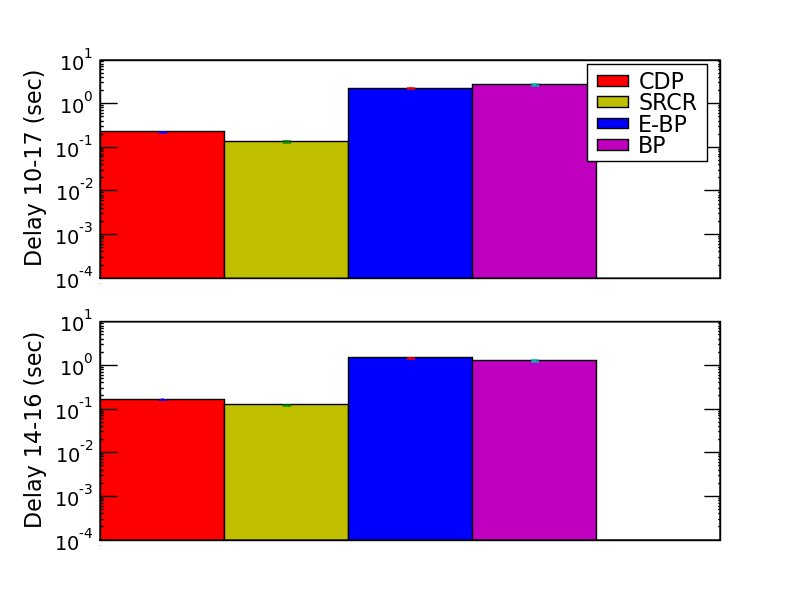}   
    \caption{Mean delay (point A)}
                 \label{delay1udp_anti} 
\end{figure}

\begin{figure}[t]
\centering
    \includegraphics[width=.40\textwidth]{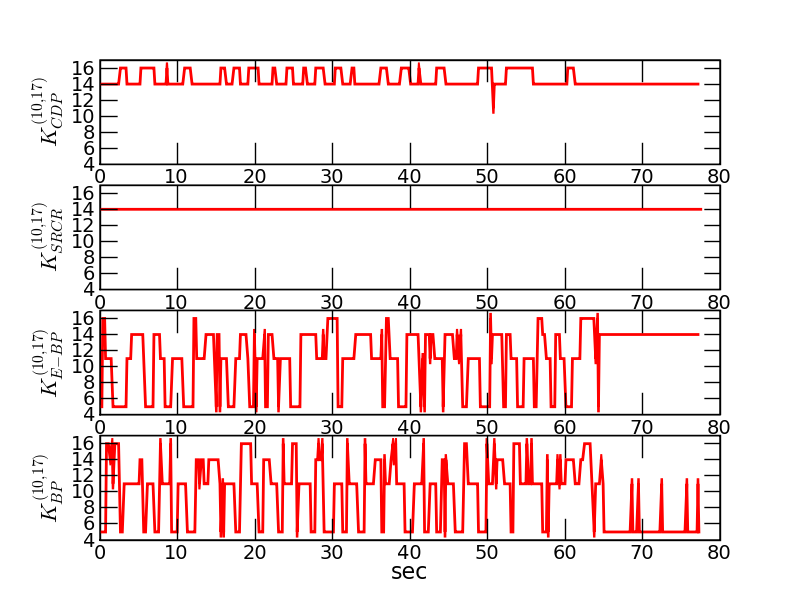}           
\caption{Routing paths taken by node 10 for the flow 10-17 (point A)}
 \label{fig:QueueNextHop_anti}
\end{figure}

\begin{figure}[!ht]
\centering
    \includegraphics[width=0.8\columnwidth]{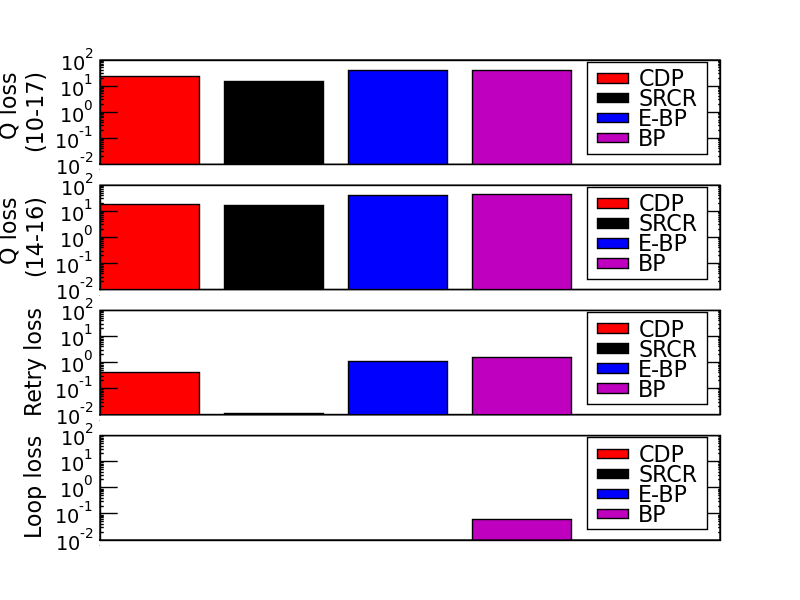}   
\caption{Loss decomposition percentage (point A)}
                 \label{loss_anti} 
\end{figure}

\subsubsection{Congestion-aware Routing: Pros  (point B)}
\label{canonical}

\begin{figure}[!ht]
\centering
{
    \includegraphics[width=0.8\columnwidth]{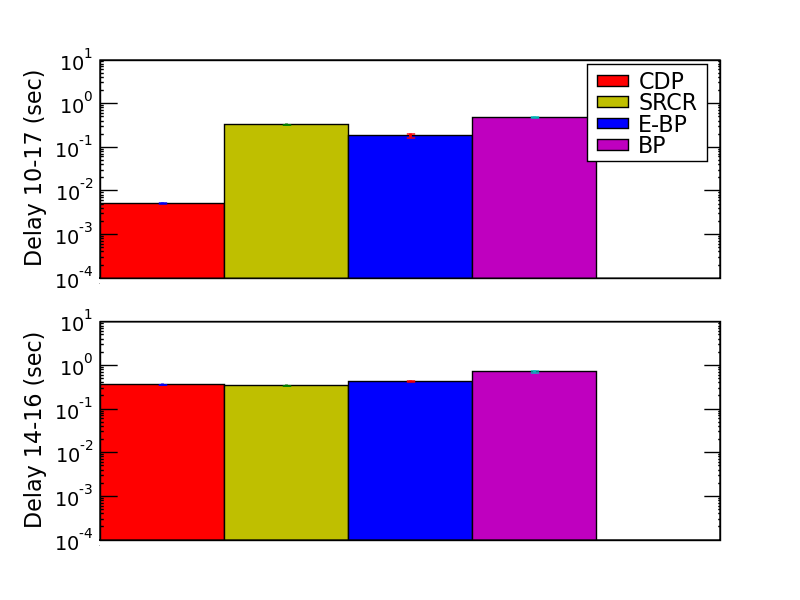}    
    }
   \caption{Mean delay (point B)}
     \label{fig:delay1udp}
\end{figure}

\begin{figure}[t]
\centering
    \includegraphics[width=.40\textwidth]{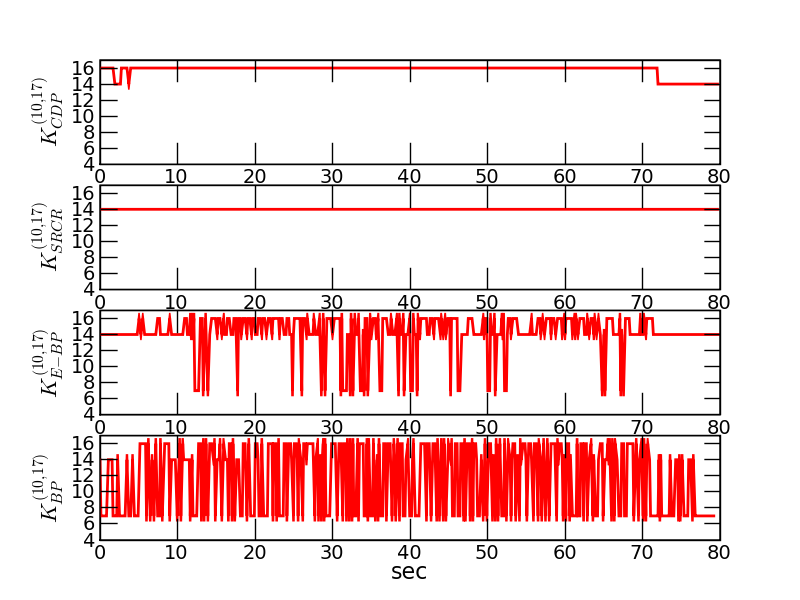}           
\caption{Routing paths taken by node 10 for the flow 10-17 (point B)}
 \label{fig:QueueNextHop}
\end{figure}
 
 We now study the flow configuration 
at point B consisting of flow 10-17 with low load and 14-16 with high load.  

 Figure \ref{fig:delay1udp} plots the end-end delay for the individual flows. The delay performance under CDP shows 
a significant improvement over the other candidate protocols  (up to a 50 fold improvement for flow 10-17, assuming the transmission time is 1 ms for each hop). 

Figure \ref{fig:QueueNextHop}  shows node 10's next hop selections. 
 We observe that SRCR persistently relies on routing via node 14, resulting in severe congestion and 
packet drops for the flow 10-17, reducing the throughput 
and increasing the delay.  BP and E-BP still forward packets to nodes 5, 7 and 11 resulting in increased interference.

  Figure \ref{fig:lossudp} shows the decomposition of sources of loss under each protocol.   In contrast to the previous example, 
Flow 10-17 shows the number of packet drops is significant for SRCR, E-BP and BP i.e.  
20\%, 25\% and 40\% respectively; while CDP packet loss is less than 1\% (mostly due to buffer overflow). 
The retry losses and loop losses for 
all protocols are negligible.

\begin{figure}[ht]
\centering
    \includegraphics[width=0.8\columnwidth]{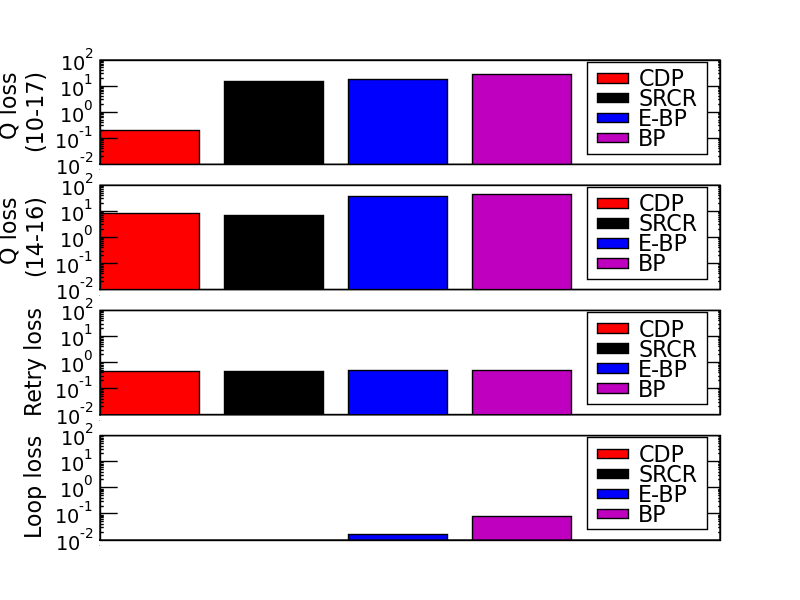} 
\caption{Loss decomposition percentage (point B)}
\label{fig:lossudp}
\end{figure}

The overall performance gain of CDP over SRCR, here suggests that there are significant gains associated with 
multi-path routing and congestion avoidance. 
However, the overall performance degradation at point A suggests that when the gains associated with multi-path and congestion-aware routing 
are small, the impact of intra-flow and inter-path interference can have a significant disadvantage for any congestion-aware routing.

\section{Modular Approach and Multi-path Diversity}
\label{roleinterfereces}
In this work, we have taken a modular approach of separating MAC from routing in designing congestion-aware routing. 
Such a constraint can be a significant limitation, however, in exploring complete capabilities of congestion and multi-path diversity. 
Unless a cross-layer approach is used, any multi-path routing algorithm suffers from a complex 
effects of interference while exploiting congestion diversity. 
In this section, we shed some light to quantify the degradations and gains in 
a modular setting by studying three simple single flow scenarios.  We claim that a modular approach 
is sufficient in real networks operating in an environment with a sufficiently high interference floor. 

\subsection{Case studies}
In this section, we consider a set of test topologies all of which contain two choices in path selection for routing. 
These class of topologies consists of a flow which can be routed using two possible choices: Path-1 or Path-2 as shown in  Figures \ref{topo_inter2}. 
 We analyse the effect of multi-path and congestion diversity using a set of randomized protocols 
where a packet at source node selects Path-1 with probability $\alpha$ (SRCR, CDP, BP and E-BP lie in this class). We then 
compare the delay under CDP with respect to $\alpha$, as $\alpha$ is varied from 0 to 1.
In the following examples, SRCR coincides with the case when $\alpha = 1$.\footnote{Note that we have not analysed the performance of BP and E-BP due to their consistently poorer performance.}

\begin{figure}[h!]
\centering
    \includegraphics[width=0.8\columnwidth]{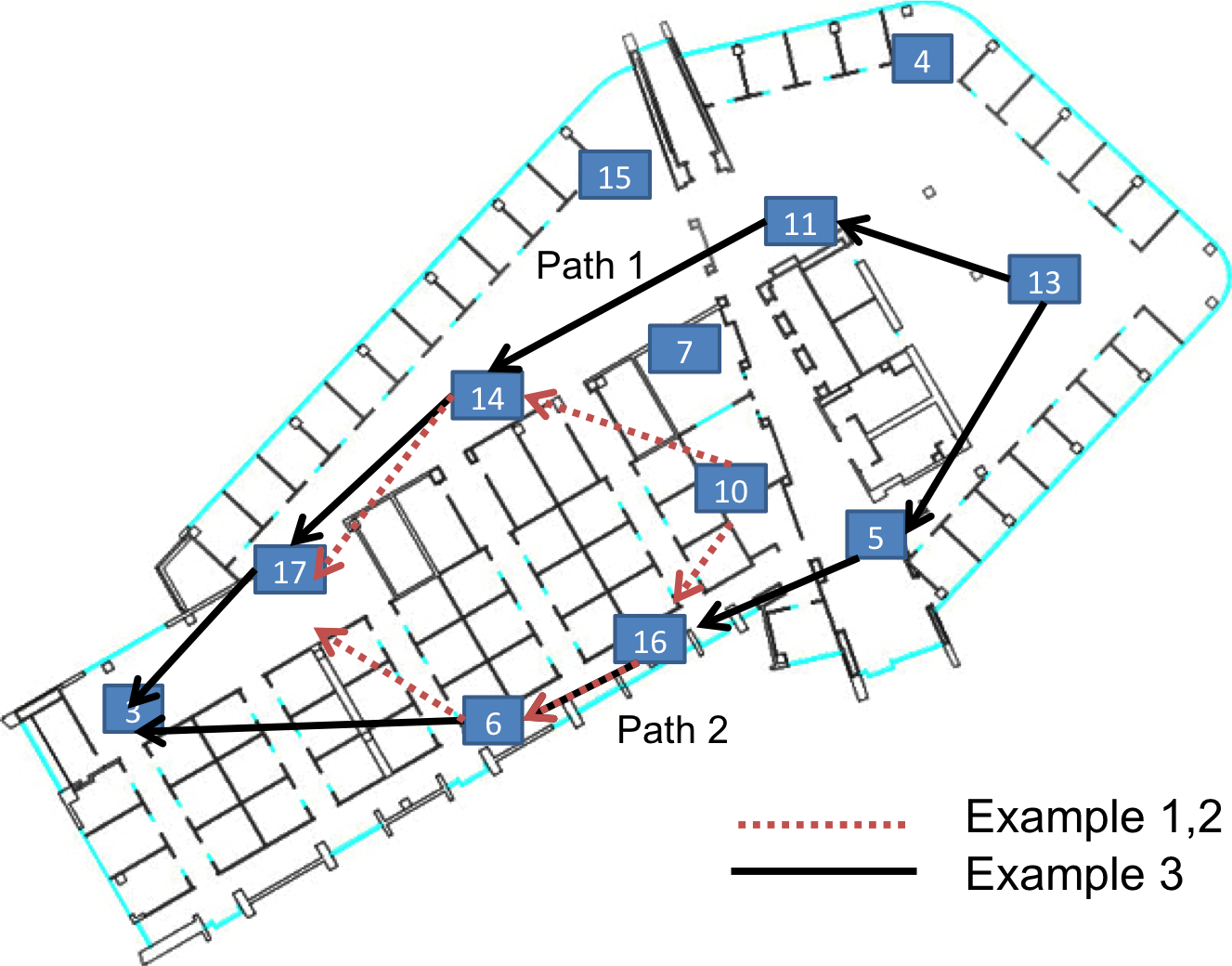}   
\caption{Topology used to study the gains of multipath diversity}
\label{topo_inter2}
\end{figure}

\subsubsection{Example 1} 
  
\begin{figure}[h]
\centering
\includegraphics[width=0.8\columnwidth]{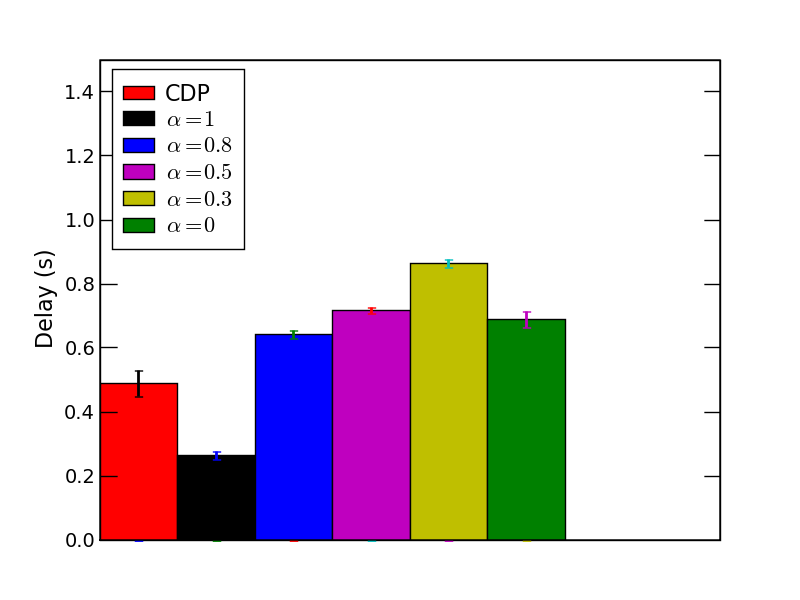}%
\caption{Delay performance as $\alpha$ is varied for the topology in Figure~\ref{topo_inter}}
\label{drain4}
\end{figure}

 We analyse the topology in Figure \ref{topo_inter2} for a high load of 4 Mbps for flow 10-17.\footnote{This topology is similar to the topology 
 for point B in Figure \ref{topo_inter}.}
Figure~\ref{drain4} shows the end-end delay performance as $\alpha$ is varied. It shows that the delay performance is poor for CDP  and  
for intermediate $\alpha$, $0 < \alpha < 1$, i.e. when diversity is utilized. The end-end delay increases significantly for $\alpha$ near 0.5. When interfering paths (Path-1 and Path-2) are used simultaneously, nodes 10, 14, 16 and 6 transmit and contend for the 
wireless channel concurrently. Thus the gain achieved by reducing the congestion at node 14 is completely erased due to the increased delay during the channel access. This analysis also suggests the possible cause of CDP's noticeable performance degradation for point A (see Figure~\ref{delay1udp_anti}).

\subsubsection{Example 2}
We consider the flow 10-17 in Figure \ref{topo_inter2} when the queue at node 14 builds up due to the low link quality of 14-17.  In practice, low link quality for 14-17 can easily occur 
when there is an obstruction between 14-17. 
Figure \ref{alphalowpower1} shows the delay performance  for the set of routing protocols as $\alpha$ is varied.
It shows that the delay is minimum for $\alpha=0$.  As link 14-17 of low quality, packets routed through node 14 may build queue at node 14 resulting in high delay. Thus, it is best to route all packets along node 16 using $\alpha=0$.

\begin{figure}[h]
\centering
\includegraphics[width=0.8\columnwidth]{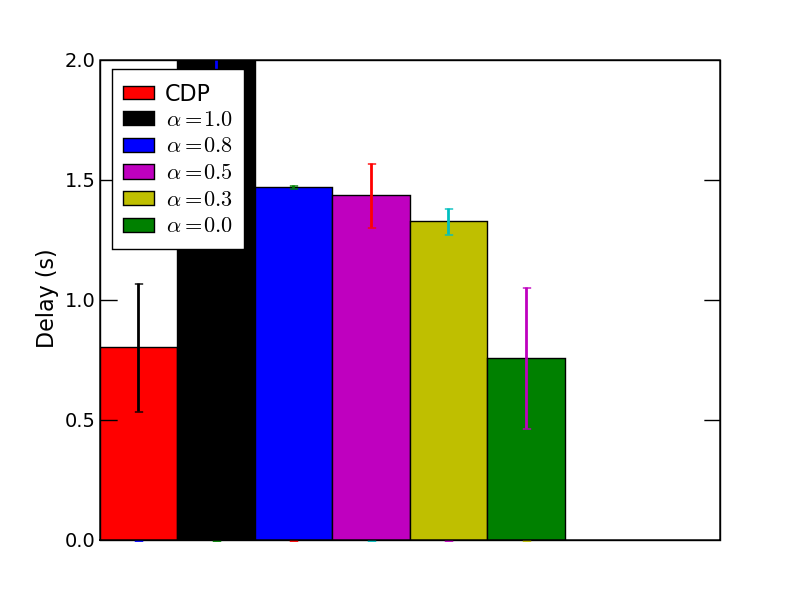}%
\caption{Delay performance as $\alpha$ is varied for case 1}
\label{alphalowpower1}
\end{figure}

\subsubsection{Example 3}

We now turn attention to another special case for flow 13-3 shown in Figure \ref{topo_inter2}, 
where the self-interference is significant on both paths
and the number of hops are high along those paths. 
Figure \ref{alpha133} shows the delay performance of the set of routing protocols as $\alpha$ is varied
for flow 13-3. 
We observe that the mean delay is minimized for $\alpha =0.5$ when congestion diversity is utilized.
It is interesting to note that the CDP is unable to follow an optimal $\alpha = 0.5$ (it tracks $\alpha = 0.38$). It is mainly due to the lag between the ideal time to switch between paths and the actual time it switches as the CDP does averaging of various quantities involved and reacts slowly.

\begin{figure}[h]
\centering
\includegraphics[width=0.8\columnwidth]{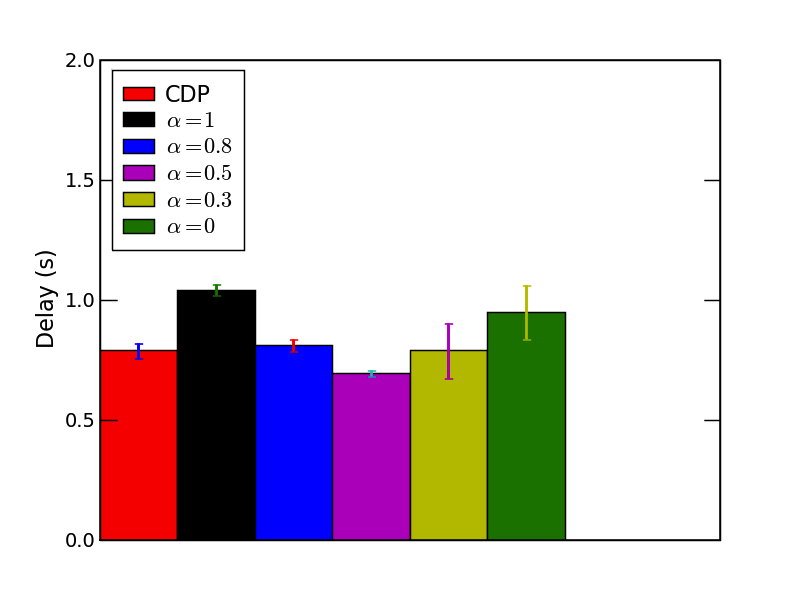}%
\caption{Delay performance as $\alpha$ is varied for case 2 }
\label{alpha133}
\end{figure}

Example 1 suggests that it is not advisable to use multi-paths when self-interfere is high; while  
examples 2 and 3 confirm the need for utilizing multi-path diversity when congestion dominates self-interference.

 This study also reveals that exploiting multi-path 
diversity is a complicated function of the intra-flow and inter-path interference in the network. Even in simple topologies, it is not clear, if multi-path diversity 
is beneficial in every scenario. 
Furthermore, it is difficult to capture the interference effects completely in any multi-path routing protocol 
unless the modular approach of network design is sacrificed. This study is left for future work. 

\subsection{Performance in high interference scenarios}
Next, we study the effectiveness of multipath routing in networks with non-negligible interference. 
In real deployments, concurrent flows already exist in the network with a sufficiently high interference floor.   
In such networks, we argue that exploiting 
congestion diversity is beneficial. In other words, we argue that in most practical scenarios, even with a modularized routing/MAC framework, the benefits of congestion-aware multi-path diversity are significant. 
 
We test the conjecture that congestion diversity gain can be dissected from the 
self interference effect in the network by equalizing the underlying interference using external interfering traffic sources. 
We repeat the experiments in Section \ref{udpexp}, where in addition to the two randomly selected heavy and long UDP flows, 
nodes in the network are  
engaged in the transmission of low intensity traffic at the rate of 10 packets/sec.

\begin{figure}[h]
\centering
\subfigure[CDF of delay differential] {
\includegraphics[width=0.8\columnwidth]{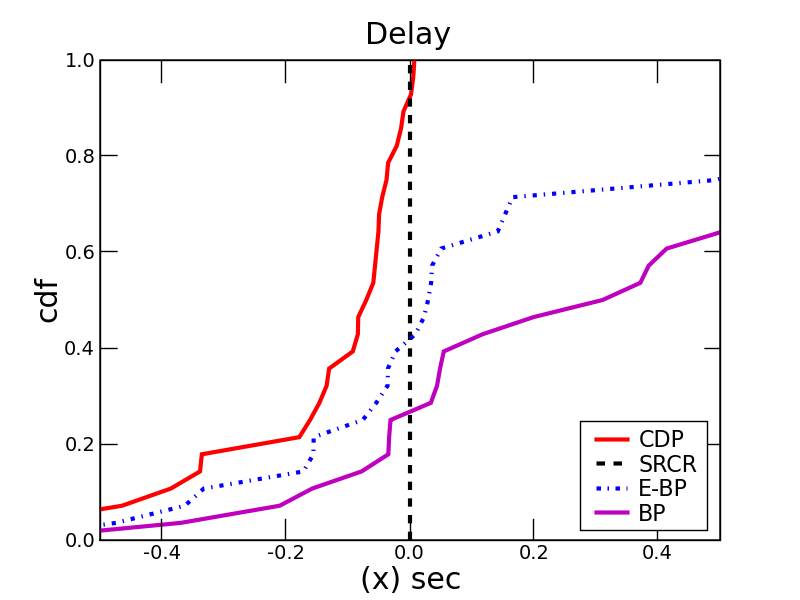}%
\label{fig:delay1}
}
\subfigure[CDF of normalized throughput] {
\includegraphics[width=0.8\columnwidth]{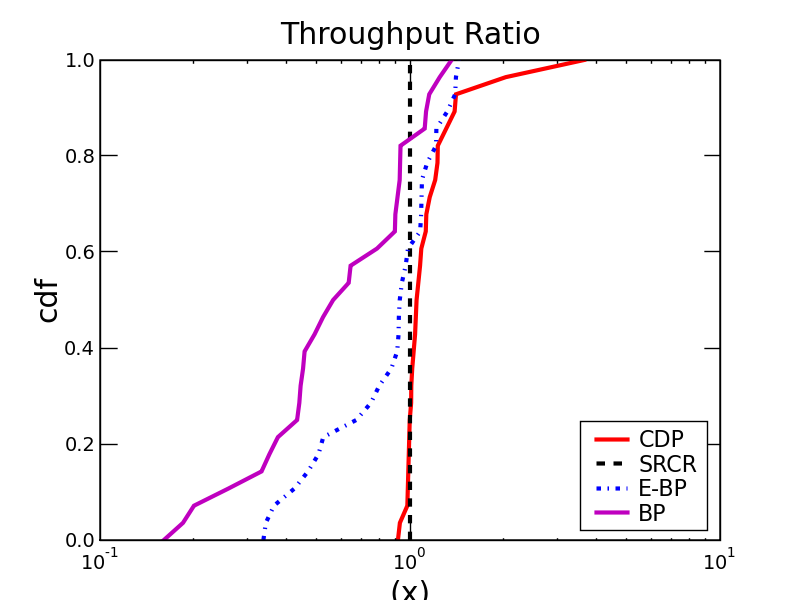}%
\label{fig:thru1}
}
\caption{Performance Results for CDP under high load with external interference}
 \end{figure}

 Figures~\ref{fig:delay1} and \ref{fig:thru1}  show further improved performance under CDP over other existing solutions. In Figure \ref{fig:delay1}, CDP shows a higher gain in terms of delay compared to that in Figure~\ref{fig:highgaindelay1}. 
Due to the existing background interference, the self-interference effect is minimized and this allows CDP to exploit the available congestion diversity.

{

\section{Discussion and Future work}
We conclude the paper with a discussion on our results, as they will guide the future directions of 
our work.
\subsection{Discussion}
\label{conclusion}
This paper presents study of a set of congestion-aware routing protocols  Backpressure (BP), Enhanced Backpressure (E-BP) and Congestion Diversity Protocol (CDP) for routing packets across a wireless multi-hop network. 
We modify the protocol stack at the routing layer to take the
congestion in the network into account. In E-BP and CDP, nodes route packets according to a rank ordering of the nodes based on a congestion measure which combines the important aspects of shortest
path routing with those of backpressure routing and are designed to alleviate the delay performance of BP. 

We evaluated these routing protocols on a real testbed of 12 nodes with end-end delay and throughput ratio as central 
metrics for comparison. 
We compared the performance of BP, E-BP  and CDP versus other routing-layer solutions, i.e.  SRCR,  under both UDP and TCP connections. We showed significant improvements for most arbitrary network set up and traffic conditions for CDP while BP and E-BP showed poorer performance.   
We also dissected network scenarios to gain insights and understand the reasons for  improvements and performance 
degradations with respect to SRCR. In the process, we shed light on the importance of a cross-layer approach in which scheduling and transport-layer congestion control enable further improvement and complimentary roles in addition to the 
congestion-aware functionality. This set of observations allow us to suggest and consider a rich set of future directions for
our work.

\subsection{Future Work}
In this work, we have taken a modular approach in which the MAC layer and the transport layer are kept intact and the routing layer is modified. 
Our results  indicate the need for the development of practical yet joint MAC, routing, and transport 
layer protocols that tackle the issues of contentions, congestion, and delay simultaneously as advocated by~\cite{Walrand},\cite{Rhee09}. This includes design of congestion-aware MAC and TCP rate control algorithm based on congestion measure $V_t$.

Given the legacy of the 802.11 MAC protocol, it might be useful to study further on the modular approach when self-interference dominates. 
Instead of changing the next hop as soon as congestion is detected,  in this case we are interested in studying a slightly modified version of CDP
to hold the route for long enough duration until we detect a sufficient congestion.

Most wireless nodes are equipped with rate selection mechanisms that attempt at optimizing the transmission rate. However, an appropriate rate selection algorithm remains as an important area of future work. 

As a side note, we would like to include a provably loop free mechanism in CDP. 
Provably loop free methods using destination sequence numbers \cite{AODV03} are slow to propagate and are unsuitable in a very dynamic system. Future work includes extending a provably loop free technique \cite{RIP-MTI} for CDP.

\end{document}